\documentclass{article}

\usepackage{arxiv}

\usepackage[utf8]{inputenc} 
\usepackage[T1]{fontenc}    
\usepackage{hyperref}       
\usepackage{url}            
\usepackage{booktabs}       
\usepackage{amsfonts}       
\usepackage{nicefrac}       
\usepackage{microtype}      
\usepackage{lipsum}
\usepackage{graphicx}
\usepackage[frozencache,cachedir=.]{minted}
\usepackage{xcolor}
\usepackage{subcaption}
\usepackage{amsmath}
\usepackage{xspace}
\usepackage{fancyhdr}
\usepackage{wrapfig}
\usepackage{enumitem}
\usepackage{mathpartir}
\usepackage{threeparttable}
\graphicspath{ {./figures/} }

\definecolor{bg}{rgb}{0.95,0.95,0.95}

\newcommand{\sysname}{{ReCopilot}\xspace}
\newcommand{\eg}{\textit{e.g.}\@\xspace}
\newcommand{\ie}{\textit{i.e.}\@\xspace}

\title{ReCopilot: Reverse Engineering Copilot in Binary Analysis}

\author{
Guoqiang Chen \quad
Huiqi Sun \quad
Daguang Liu \quad
Zhiqi Wang \quad
Qiang Wang \quad \\
\textbf{Bin Yin} \quad
\textbf{Lu Liu} \quad
\textbf{Lingyun Ying} \\
\texttt{\{guoqiangchen, sunhuiqi, liudaguang, wangzhiqi, wangqiang10,} \\
\texttt{binyin, liulu01, yinglingyun\}@qianxin.com} \\
QI-ANXIN Technology Research Institute, Beijing, China
}

\begin{document}
\maketitle

\begin{abstract}
Binary analysis plays a pivotal role in security domains such as malware detection and vulnerability discovery, yet it remains labor-intensive and heavily reliant on expert knowledge. 
General-purpose large language models (LLMs) perform well in programming analysis on source code, while binary-specific LLMs are underexplored. 
In this work, we present \sysname, an expert LLM designed for binary analysis tasks. \sysname integrates binary code knowledge through a meticulously constructed dataset, encompassing continue pretraining (CPT), supervised fine-tuning (SFT), and direct preference optimization (DPO) stages. It leverages variable data flow and call graph to enhance context awareness and employs test-time scaling to improve reasoning capabilities. Evaluations on a comprehensive binary analysis benchmark demonstrate that \sysname achieves state-of-the-art performance in tasks such as function name recovery and variable type inference on the decompiled pseudo code, outperforming both existing tools and LLMs by 13\%. Our findings highlight the effectiveness of domain-specific training and context enhancement, while also revealing challenges in building super long chain-of-thought. \sysname represents a significant step toward automating binary analysis with interpretable and scalable AI assistance in this domain.
\end{abstract}


\section{Introduction}
\label{sec:introduction}
Binary analysis plays a vital role in cybersecurity, empowering security professionals to uncover vulnerabilities, detect backdoors, and identify malware effectively. However, the lack of symbolic information in stripped binaries poses a significant challenge in understanding the binary code and performing analysis. Traditional decompilers, such as IDA Pro~\cite{IDA} and Ghidra~\cite{Ghidra}, are powerful tools for reverse engineering and could generate C-like pseudo code, but they often struggle with predicting missing symbols, such as variable name and type, which are crucial for understanding the code's functionality. The decompiled pseudo code without source-level symbols is still difficult to read and understand, making it challenging for security analysts to reach the functional semantics efficiently.

General-purpose large language models (LLMs) have shown remarkable capabilities in various programming tasks, including code generation~\cite{Leung2023ase, Pearce2022sp, deepseek-coder-v2-2024} and bug fixing~\cite{jiangICSE2023, gao2023far, Kang2023ICSE}. A recent investigation study~\cite{LLM-code-eval-openai2025} has demonstrated that LLMs performed close to top-tier human competitors in competitive programming challenges. Considering the success of LLMs in the source-code domain, it is intuitive to explore their potential in binary analysis. The recent studies~\cite{llm4decompile-2024emnlp, ReSym-2024ccs, varbert-2024sp} have shown that LLMs can be effectively utilized to inferring symbols or reconstructing source code to assist the analysts. 
However, these efforts are still focusing on single task, such as decompilation~\cite{llm4decompile-2024emnlp} or variable name/type prediction~\cite{ReSym-2024ccs, varbert-2024sp}, leading to their lack of application potential in practice scenarios.

In this paper, we present an expert LLM, named \sysname, which is designed to be a \textit{R}everse \textit{E}ngineering \textit{COPILOT} in binary analysis with multiple tasks supported, including decompilation, function name recovery, variable name/type recovery, struct recovery, and binary code summarization, etc. \sysname is trained on domain-specific data and tuned for equipping test-time scaling capability, which allows it to take a deep reasoning to improve the accuracy of the final prediction. 
To build the model, we constructed three datasets to launch different training stages, including continue pretraining (CPT), supervised fine-tuning (SFT), and direct preference optimization (DPO).
We also emitted a static program analysis on decompiled pseudo code to enhance the model's understanding of the code context and variable usages.
Beyond the model building, we also developed a comprehensive benchmark to assess the performance of \sysname on various binary analysis tasks. The results demonstrated that \sysname achieved comparable performance to advanced general LLMs (DeepSeek-V3 671B) with a much smaller model size (7B), and outperforms the existing domain models by 13\% averaged across tasks.
To facilitate the security community, we have made the demo of \sysname publicly accessible to promote further research in this area\footnote{\url{https://tqgpt.qianxin.com/recopilot}}.

\section{Background}
\label{sec:background}
The source code is generally compiled into low-level machine code and assembled into executable files or runtime libraries, called binaries, which can be loaded into memory and executed by the CPU. 
The modern decompiler could lift machine code in bit stream into a high-level representation, \ie, C-like pseudo code as shown in \autoref{fig:pcode-example-pcode}. Unfortunately, the debug symbols are often stripped before releasing the binaries for various purposes such as reducing file size and hiding functionality.
The missing symbolic information, such as function names, variable names, and type definitions, makes it difficult to analyze the binaries. The traditional decompilation tools struggle with recovering these information, instead using placeholders (\eg, \texttt{a1} and \texttt{v1}) in the pseudo code.

\begin{figure}[htbp]
    \centering
    \begin{minipage}[t]{0.48\textwidth} 
        \begin{minted}[
            frame=lines, % 添加边框
            framesep=2mm, % 边框与代码间距
            baselinestretch=1.2, % 行间距
            bgcolor=bg, % 背景颜色
            fontsize=\scriptsize, % 字体大小
            linenos, % 显示行号
            numbersep=2pt,
            breaklines % 自动换行
        ]{c}
void __fastcall sub_1909(__int64 a1, __int64 a2, unsigned __int64 a3)
{
  _OWORD *v3; // rcx
  unsigned __int64 i; // rbx
  _OWORD *v7; // rbp
  __int64 j; // rax
  __int64 v9; // rdi

  v3 = (_OWORD *)(a1 + 176);
  for ( i = 0LL; ; i += 16LL )
  {
    v7 = (_OWORD *)(a2 + i);
    if ( i >= a3 )
      break;
    for ( j = 0LL; j != 16; ++j )
      *((_BYTE *)v7 + j) ^= *((_BYTE *)v3 + j);
    v9 = a2 + i;
    sub_14F7(v9, a1);
    v3 = v7;
  }
  *(_OWORD *)(a1 + 176) = *v3;
}
        \end{minted}
        \subcaption{Pseudo Code}
        \label{fig:pcode-example-pcode}
    \end{minipage}
    \hfill 
    \begin{minipage}[t]{0.48\textwidth} 
        \begin{minted}[
            frame=lines,
            framesep=2mm,
            baselinestretch=1.2,
            bgcolor=bg,
            fontsize=\scriptsize,
            linenos,
            numbersep=2pt,
            breaklines
        ]{c}
#define AES_keyExpSize 176
#define AES_BLOCKLEN 16 
struct AES_ctx
{
  uint8_t RoundKey[AES_keyExpSize];
  uint8_t Iv[AES_BLOCKLEN];
};
void AES_CBC_encrypt_buffer(struct AES_ctx *ctx, uint8_t* buf, size_t length)
{
  size_t i;
  uint8_t *Iv = ctx->Iv;
  for (i = 0; i < length; i += AES_BLOCKLEN)
  {
    XorWithIv(buf, Iv);
    Cipher((state_t*)buf, ctx->RoundKey);
    Iv = buf;
    buf += AES_BLOCKLEN;
  }
  /* store Iv in ctx for next call */
  memcpy(ctx->Iv, Iv, AES_BLOCKLEN);
}
        \end{minted}
        \vspace{-0.1cm}
        \subcaption{Source Code}
        \label{fig:pcode-example-scode}
    \end{minipage}
    \caption{An example of decompiled pseudo code and the corresponding source code from a real-world \texttt{AES} implementation \cite{tiny-aes-c}.}
    \label{fig:pcode-example}
\end{figure}

\autoref{fig:pcode-example} shows an example of pseudo code decompiled by IDA Pro and its corresponding source code, which is an AES encrypt function from an open source repository~\cite{tiny-aes-c}. 
It is difficult to identify the high-level semantics of the pseudo code function even though it exhibits a syntactic structure similar to the source code. 
As for the source code, we can easily make a determination even with just the function name, but unfortunately, these symbols are erased from the stripped binaries.
On the other hand, binary analysis tasks, like vulnerability detection, often require a deep understanding of the data structure of a variable, which is often lost in the decompiled code.
Based on the source code, it is easy to tell that the first parameter of the function shown in \autoref{fig:pcode-example} accepts a struct variable of type \texttt{AES\_ctx}, but such conclusion is not straightforward in the decompiled code. We can only infer from \texttt{L9} and \texttt{L16} that the position at an offset of 176 from variable \texttt{a1} is a 16-byte length data. However, to determine the specific structure of the first 176 bytes of \texttt{a1}, more information needs to be gathered by analyzing both the caller and callee of \texttt{sub\_1909}. 
In general, it is much harder to understand the functionality of the decompiled pseudo code, since it has no meaningful symbol. 

In recent years, LLMs have shown great potential in understanding, generating, and analyzing programs, providing substantial assistance to programming tasks in the source code.
EvalPlus~\cite{evalplus, evalperf} hosts a leaderboard for evaluating the performance of LLMs on code synthesis. As of the completion of this paper, EvalPlus with HumanEval~\cite{humaneval-chen2021} dataset have been almost fully passed by the top models, with o1-preview achieving a score of 96.3.
SWE-bench~\cite{SWE-bench-2024} is an evaluation framework consisting of real GitHub issues, designed to measure the performance of LLMs in resolving codebase defects. According to the SWE-bench leaderboard, the SOTA method driven by Claude 3.7 Sonnet has successfully resolved 33\% of the issues in 2025, highlighting a significant and rapid improvement compared to the 2\% achieved by the best-performing model in 2023.

Given the success of LLMs in source code, it is intuitive to prompt general LLMs to perform binary code analysis, such as vulnerability detection and symbolic information recovery. There are several investigation studies that show the potential and limitation of these LLMs in binary analysis tasks. 
BinSum~\cite{BinSum-jin2023} present a large-scale assessment of LLMs on binary code summarization, and show that they can generate high-quality summaries for binary functions, however, the performance is dropping significantly facing stripped binaries. Another work~\cite{howfarbin-shang2024} also evaluate the LLMs in binary code understanding with function name prediction and binary code summarization tasks, and drew similar conclusions.
DeBinVul~\cite{DeBinVul-manuel2024} employs vulnerability detection, classification, and summarization tasks for evaluation, revealing the sub-optimal performance of general models compared to the fine-tuned domain models.
In general, we consider binary analysis as a domain-specific problem, where general-purpose LLMs, without specialized training, still struggle to reach the level of human experts. In other words, developing expert models presents a promising direction for integrating AI technologies into the field of binary program analysis.

\noindent{\textbf{Related Work.}} There are several attempts to predict the symbolic information of binaries using neural-based techniques. The following research efforts~\cite{DEBIN-CCS2018, NERO-OOPSLA2020, NFRE-2021ISSTA, SYMLM-CCS2022, NER-2023pst, XFL-2023sp, Epitome-2024FSE} are designed to predict the source-level names of binary functions to facilitate understanding by analysts.
There are also several studies that focus on predicting the variable names~\cite{DIRTY-2022usenix, DIRECT-2021ACL, DIRE-ASE2019, HEXT5-2023ase, varbert-2024sp} and types~\cite{DIRTY-2022usenix,stateformer-2021fse, TypeSqueezer-2023ccs, TYGR-2024usenix, TypeFSL-2024ase, ReSym-2024ccs} in the decompiled pseudo code by developing neural network-based approaches, and some of them, such as TYGR~\cite{TYGR-2024usenix} and ReSym~\cite{ReSym-2024ccs}, attempt to predict the memory structure of complex variables (\eg, struct). Meanwhile, others make efforts to fine-tune models to learn code semantics and generate comprehensive summaries~\cite{BinT5-2023saner, HEXT5-2023ase} even source code~\cite{HEXT5-2023ase, llm4decompile-2024emnlp} for binary functions. 
All of these studies have achieved promising results in their respective evaluations and offered valuable insights for subsequent research.
However, they are limited to addressing only one specific task, rendering their methods insufficient for meeting the diverse demands of real-world binary analysis.
On the other hand, only a few studies, like LLM4Decompile~\cite{llm4decompile-2024emnlp}, attempt to leverage LLMs for binary analysis tasks, which leaves a lot of room for our research.


\noindent{\textbf{Challenges.}} The easiest approach is to prompt existing LLMs to perform binary analysis tasks. As mentioned above, however, previous evaluations~\cite{BinSum-jin2023, howfarbin-shang2024, DeBinVul-manuel2024} have demonstrated that it yields limited performance. Therefore, we here aim to train an expert LLM to support the most important and common tasks in binary analysis practice. To this end, we need to address several challenges:

1) Lacking publicly available datasets, we need to collect large-scale and fine-grained domain data, specifically align data between stripped binaries and source code.

2) Considering the offline analysis requirements in practice, we expect a small LLM to support local inference, which should be laptop-deployable. This motivates us to train the model carefully to prevent overfitting while ensuring good performance.

3) The context limitation of LLMs constrains us to analyze only a portion of a large binary program at one time. The previous methods typically make the prediction based on a single binary function, resulting in providing insufficient information. Therefore, we aim to enable the model to utilize context information to improve analysis, which requires additional efforts in context building.

\section{Methodology}
\label{sec:methodology}
\subsection{Overview}

\begin{figure}[htbp]
    \centering
    \includegraphics[width=0.9\textwidth]{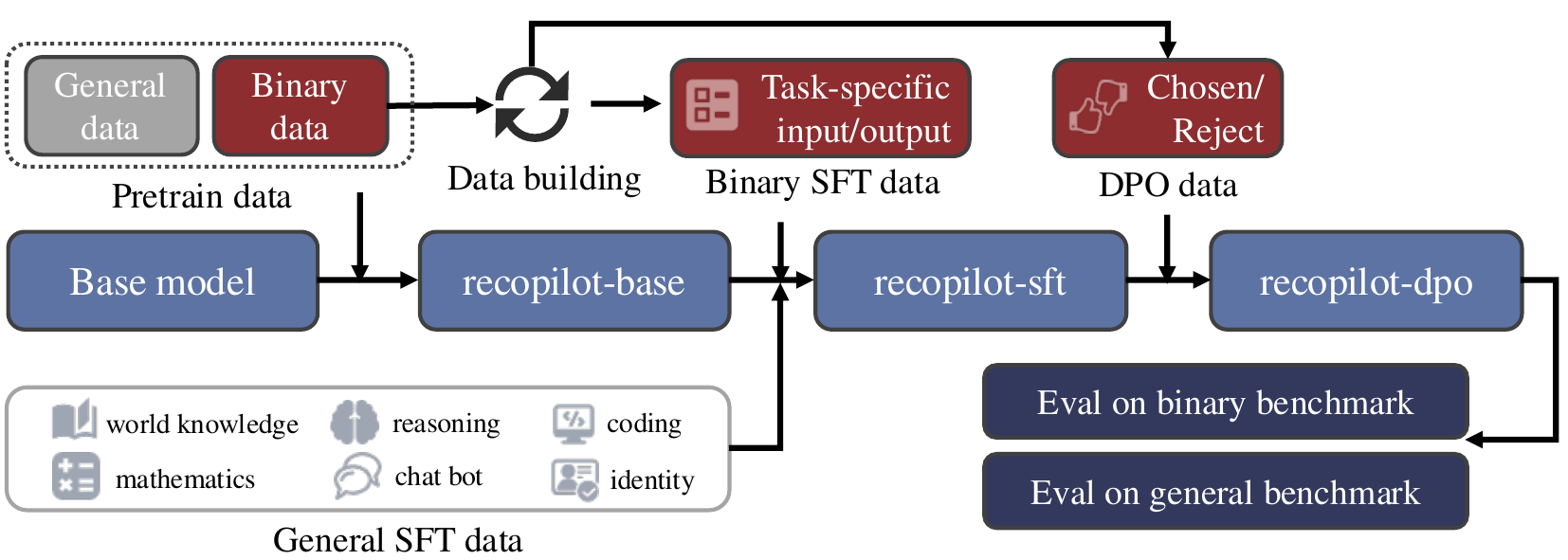}
    \caption{An overview of the \sysname model building.}
    \label{fig:overview}
\end{figure}



\autoref{fig:overview} shows the overview of our model building process that mainly consists of three training stages. Specifically, we first conduct continued pretraining (CPT) on a pretrained base model to learn domain language and knowledge. 
Then, a supervised fine-tuning (SFT) is performed with a mixture SFT dataset to empower the model with reasoning ability and adapt it to the downstream tasks. 
Finally, we employ the direct preference optimization (DPO) to further improve the model's performance in the format-following. We will introduce the details of each stage in \S\ref{sec:training-strategy}.

We have built corresponding datasets from scratch to launch these trainings. Several pipelines are firstly designed to collect the raw binary dataset, which is then filtered and sanitized to build the final training datasets. To build a task-specific SFT dataset for binary analysis, we devise a generator-discriminator framework to automatically generate domain examples with chain of thought. In general, we build a large-scale pretrain dataset with 60B tokens, a mixture SFT dataset with 1.7B tokens, and a DPO dataset with 2.4K examples, and they will be detailed in \S\ref{sec:dataset-building}. 

We also propose a context enhancement method to improve the prompt building for the target binary function. 
The context enhancement employs static program analysis to build function context, call chains, and variable data flow, which are organized into the model input to provide as comprehensive a view as possible.
We will present the whole prompt specification and more building details in \S\ref{sec:context-enhancement}.

\subsection{Training Strategy}
\label{sec:training-strategy}

\noindent\textbf{CPT.} Pretraining is generally used to acquire foundational knowledge and learn language features from large-scale data, enabling efficient adaptation to diverse downstream tasks through transfer learning. General-purpose LLMs~\cite{llama3-2024, deepseekv3-2025, qwen25-2025} are pretrained on wide-ranging corpora, such as books, Wikipedia, and web pages, which rarely contain binary code (\ie, decompiled pseudo code). And the publicly exposed pseudo code often lacks corresponding source code information, indicating a low data quality. These facts suggest that the general LLMs are underfitting to the binary domain. 

To better model pseudo code, we conducted continued pretraining (CPT) on a base LLM using a customized dataset.
Meanwhile, we took further efforts to make the model learn the mappings between binary code, source code, and natural language. Overall, our CPT injected domain knowledge into the model and improves its understanding of binary code, which is crucial for the subsequent fine-tuning on downstream tasks.

\noindent\textbf{SFT.} We employed supervised fine-tuning (SFT) to adapt the pretrained model (\ie, recopilot-base in \autoref{fig:overview}) to the pre-defined tasks for binary analysis. 
SFT enables the model to follow user instructions and generate results in specific formats, which is essential for our downstream applications. 
Furthermore, we consider binary analysis tasks generally are reasoning-intensive, like solving math problems, indicating that the model needs to learn how to reason about code semantics and generate accurate results.
Inspired by OpenAI-o1~\cite{openaio1-2024} and DeepSeek-R1~\cite{deepseekr1-2025}, we plan to equip our model with the test-time scaling ability. 
To this end, we propose a generator-discriminator framework (detailed in \S\ref{sec:generator-discriminator}) to automatically generate SFT data with Chain-of-Thought (CoT), enabling us to fine-tune our model to take a deep thinking before giving the answer. 
Finally, our recopilot-sft model accepts a prompt built from a decompiled binary, and generates a reasoning process for the user-specific task following with the final prediction in JSON format, which is easy to parse and apply into decompilation tools.

\noindent\textbf{DPO.} We further deployed direct preference optimization (DPO)~\cite{dpo-2023} to improve the recopilot-sft's performance in format-following and reasoning consistency.
Format errors directly affect the usability of the model, and rigorous logic is crucial for generating correct reasoning. 
Specifically, DPO is a reinforcement learning method that optimizes the model's output by directly learning from user preferences, involving data pairs that consist of chosen and rejected responses for the same prompt.
In contrast to traditional RLHF~\cite{RLHF-2022} methods, it does not need human feedback and a complex reward function, and allows us launch the training efficiently on limited computation resources.

\subsection{Dataset Building}
\label{sec:dataset-building}

\subsubsection{Raw Dataset Collection}
\label{sec:raw-data-collect}

To build a large-scale, fine-grained raw dataset of binary functions, we start with numerous stripped binaries and source code packages. In practice, we identified different data sources and designed several automatic pipelines to collect the data. Specifically, we have built the following pipelines:

\begin{itemize}[left=1em]
\item \noindent\textbf{1) Compile from Scratch:} We pull source code from the online package repositories, such as Archlinux~\cite{archlinuxpackages}, which host a large number of open-source software packages and organize them in a well-structured manner, enabling batch compilation with specific arguments. 
We further utilized a compiler wrapper to precisely control compilation options to building stripped binaries and corresponding debug information. 
Although the compilation process often fails for various reasons, such as missing dependency libraries, we still collected a large amount of data from this pipeline, specifically $\approx$60 million binary functions.

\item \noindent\textbf{2) Off-the-shelf Software Artifacts:} To enrich the dataset, we also collected binaries from existing software artifacts repositories, such as the Ubuntu~\cite{ubuntupackages} and Debian~\cite{debianpackages} package sources.
One triplet is clustered by the project name and version, and consists of a release package, a debug symbol package, and a source code package. 
These software packages are built from various compilation environments far beyond the compilers and options we used in the ``Compile from Scratch'' pipeline, which helps us to cover a wider range of real-world programs.

\item \noindent\textbf{3) CompileAgent:} To further include ad-hoc projects, we employed a recent work, CompileAgent~\cite{CompileAgent-2025}, to automatically building project from source code. CompileAgent is a LLM-agent driven framework that can take over compilation from a specific repository URL or a local codebase and handle the possible errors during the whole process. It mitigates the limitation of handling compilation errors automatically in other pipelines, providing us with critically needed projects without more manual efforts.

\end{itemize}

We obtained large-scale binary files and corresponding source code packages from these pipelines. To connect the binary functions to their source code counterparts, we leverage the debug information that contains the file path and line number for each function definition. The binary functions are decompiled by the modern decompiler IDA Pro~\cite{IDA}, and the source function are extracted by the programming parser tree-sitter~\cite{tree-sitter}. For the current version, we mainly focus on C/C++ binaries and source code. 

\noindent\textbf{Sanitization and Deduplication.} Data noise could bias the model and lead to overfitting. To improve the data quality, we performed a series of sanitization and deduplication steps. 
First, we removed the binary function with too short/long length, which could contain insufficient or excessive information to prevent effective learning. 
The thunk functions only consist of jump instructions to forward calls to other functions, and the auxiliary functions (\eg, \texttt{register\_tm\_clones}) are generated by the compiler for assisting the program execution, we thus filtered out these functions to reduce noise. 
Moreover, we also removed the functions with no source code found, \ie, missing the ground truth, which mostly correspond to third-party library code introduced in the binaries.
On the other hand, there are many similar pseudo code functions in the dataset, which could be due to the reusing code across projects. To reduce the redundancy, we applied the MinHash~\cite{minhash} algorithm to perform function-level deduplication on our raw dataset. MinHash is a locality-sensitive hashing algorithm that can efficiently identify similar items in large datasets. 

In general, our raw dataset consists of 101M binary functions collected from 11K projects, which is far beyond the 100K functions collected in previous work LLM4Decompile~\cite{llm4decompile-2024emnlp}, and the detailed statistics are shown in \autoref{tab:raw-dataset-statistics}:

\begin{table}[htbp]
    \caption{Statistics for our raw dataset in the binary domain.}
    \centering
    \label{tab:raw-dataset-statistics}
    \begin{tabular}{@{}lrrr@{}}
        \toprule
        Pipeline                         & \# Project & \# Binary & \# Function \\ \midrule
        Compile from Scratch             & 4,350          & 506,138        & 59,927,927      \\
        Off-the-shelf Software Artifacts & 7,021          & 340,029        & 40,750,991      \\
        CompileAgent                     & 101            & 9,733           & 880,414       \\
        \textbf{In Total}     & \textbf{11,472}          & \textbf{855,900}        & \textbf{101,559,332}      \\ \bottomrule
    \end{tabular}
\end{table}

\subsubsection{Pretraining Dataset}
\label{sec:pretraining-dataset}

\begin{figure}[htbp]
    \centering
    \includegraphics[width=0.9\textwidth]{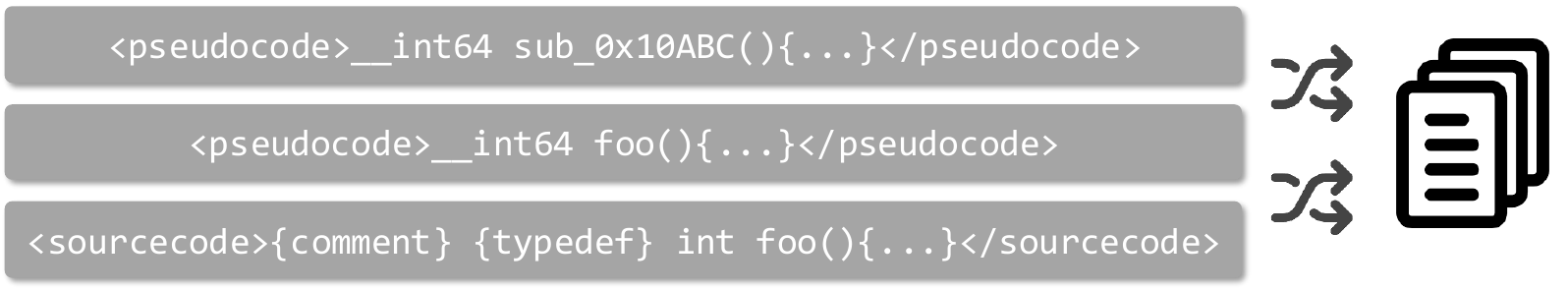}
    \caption{An example for demonstrating the pretraining data format and inner-shuffling.}
    \label{fig:pt-dataset-example}
\end{figure}

We sampled a large-scale domain pretraining dataset from the raw dataset, which is used to learn the data representation and build connections across binary, source, and natural language. To this end, as shown in \autoref{fig:pt-dataset-example}, each sample is constructed by a stripped binary function in pseudo code, the corresponding source code, and a comment in natural language. We also included the decompiled pseudo code with debug symbols in each sample, facilitating the model to learn the binary code better. Meanwhile, we involved the definitions for complex types (\ie, struct and enum) in the source code part. LLM learns the material in pretraining through the next-token prediction task, which means learning to predict subsequent tokens from preceding token sequences in a piece of data. In order to build a bidirectional mapping of different data modals rather one way, we took inner-shuffling to disorder three segments in one single data. 

To prevent our model from overfitting on binary data, we further collected general text and code data from the open-source datasets. Specifically, we randomly sampled C/C++/Python/Rust/Go/Shell code from The Stack v2~\cite{stackv2-2024} and RedPajama~\cite{redpajama-2024nips} datasets. Regarding the natural language data, we sampled from the following high-quality datasets: wikipedia~\cite{wikidump}, stackoverflow-posts~\cite{stackoverflow-posts}, and security-paper-datasets~\cite{security-paper-datasets}, where only the English and Chinese texts are selected, and we prioritize reverse-engineering related documents for sampling.

Being focused on building an expert LLM, we proactively make our binary code data dominant in the final pretraining dataset.
With references to previous work~\cite{CMR-2024acl, qwen25coder-2024}, we set the mixture ratio of binary/code/text data to 60:25:15. In summary, our final pretraining dataset contains 36B tokens, including 21B tokens from binary domain, 10B tokens from general code, and 5B tokens from natural language. The detailed statistics are shown in \autoref{tab:pretraining-dataset-statistics}.

\begin{table}[htbp]
\centering
\caption{Statistics of our pretraining (PT) dataset.}
\label{tab:pretraining-dataset-statistics}
\begin{tabular}{@{}lcrr@{}}
\toprule
Data Source          & Domain   & \# Samples          & \# Tokens(B)   \\ \midrule
the-stack-v2~\cite{stackv2-2024} & code           & 4,805,445           & 5.58           \\
RedPajama~\cite{redpajama-2024nips}  & code             & 4,343,832           & 4.08           \\
wikipedia~\cite{wikidump}        & text       & 3,593,584           & 2.04           \\
security-paper-datasets~\cite{security-paper-datasets} & text  & 428,155             & 0.53           \\
stackoverflow-posts~\cite{stackoverflow-posts}  & text   & 3,964,004           & 2.48           \\
binary-raw-dataset    & binary  & 5,733,356           & 21.18          \\
\textbf{In Total}   &  \textit{/}    & \textbf{20,233,101} & \textbf{35.83} \\ \bottomrule
\end{tabular}
\end{table}

\subsubsection{Generator-Discriminator Framework for SFT Dataset}
\label{sec:generator-discriminator}

To build the binary analysis SFT dataset, we first identified 14 tasks that are commonly used in the field and the most helpful for the participants in reverse engineering. 
As mentioned in \S\ref{sec:background}, the related works are mainly focused on function name recovery, variable name/type prediction, and binary code summarization, which are generally used to assist binary code understanding and considered as the most important tasks in this field. Based on these primitive tasks, we further defined the following 14 specific tasks to fine-tune our base model. The following is the task tags and descriptions: 

\begin{itemize}[left=1em]
    \item \textbf{Function Name Recovery:} \texttt{<funcname>
    } Given a function in decompiled pseudo code, recovery the function name in the source-code level.
    \item \textbf{Function Signature Recovery:} \texttt{<signature>
    } Given a function in decompiled pseudo code, recovery the signature (i.e., function definition) into the source-code level.
    \item \textbf{Variables Recovery:} \texttt{<vars>
    } Given a function in decompiled pseudo code, recovery the variables into the source-code level, including the variables' types (including structs) and names.
    \item \textbf{Arguments Recovery:} \texttt{<args>
    } Given a function in decompiled pseudo code, recovery the arguments into the source-code level, including the arguments' types (including structs) and names.
    \item \textbf{Variable Recovery:} \texttt{<var:var\_name>
    } Given a function in decompiled pseudo code, recovery the specific variable into the source-code level, including the variable' type (including struct) and name.
    \item \textbf{Argument Recovery:} \texttt{<arg:arg\_name>
    } Given a function in decompiled pseudo code, recovery the specific argument into the source-code level, including the argument' type (including struct) and name.
    \item \textbf{Algorithm Identification:} \texttt{<algorithm>
    } Given a function in decompiled pseudo code, identify whether this function is a particular algorithm, or part of its implementation.
    \item \textbf{Category Identification:} \texttt{<category>
    } Given a function in decompiled pseudo code, identify the functionality category of this function.
    \item \textbf{Brief Summary:} \texttt{<summary-brief-en>, <summary-brief-cn>
    } Given a function in decompiled pseudo code, generate a brief summary (1-2 sentences maximum) of the function in natural language.
    \item \textbf{Detailed Summary:} \texttt{<summary-en>, <summary-cn>
    } Generate a detailed summary in English describing the function's purpose, arguments, return value, functionality category, and possible algorithm, as well as inline comments if should have.
    \item \textbf{Binary Function Analysis:} \texttt{<func-analysis>
    } Given a function in decompiled pseudo code, conduct a detailed analysis to recovery the following information: return type, function name, arguments, variables, detailed comments, functionality category and possible algorithm.
    \item \textbf{Decompilation:} \texttt{<decompilation>
    } Given a function in decompiled pseudo code, improve the pseudo code, make it closer to source code and more understandable.
\end{itemize}

Notably, the binary function analysis task tagged by \texttt{<func-analysis>} is an overall analysis for the binary function user asked and almost covers all of the other tasks. However, we still involve multiple independent tasks to fine-tune our model rather only the one, primarily due to the transferability between tasks, where different tasks exhibit mutual benefits. For example, improvements in type inference can potentially enhance the model performance in the semantic understanding task and vice versa.

\begin{wrapfigure}{r}{0.4\textwidth}
    \centering
    \includegraphics[width=0.39\textwidth]{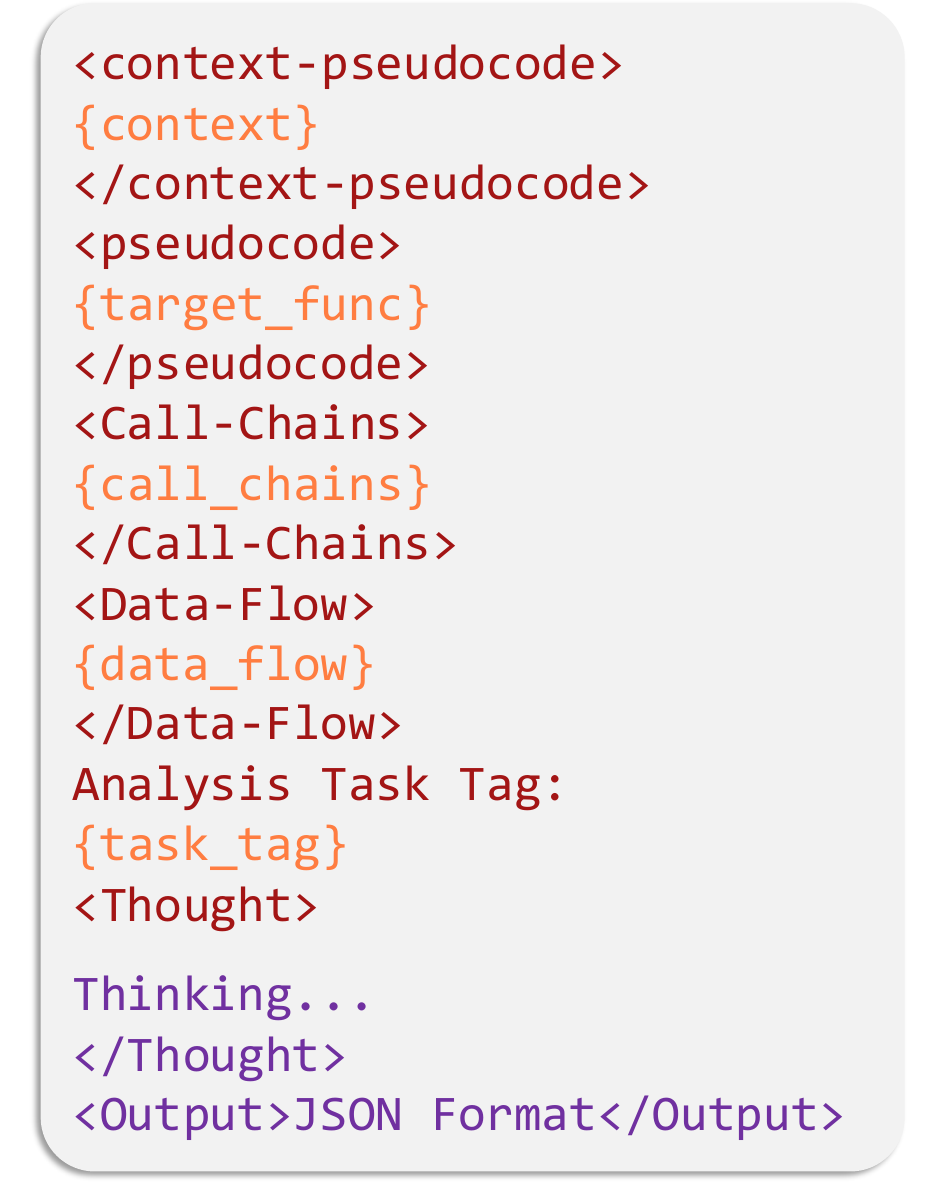}
    \caption{The \textcolor[HTML]{95261F}{input}-\textcolor[HTML]{68349A}{output} template designed for binary analysis tasks in \sysname.}
    \label{fig:sft-template}
\end{wrapfigure}

We carefully defined a general input-output template for the binary analysis tasks, as shown in \autoref{fig:sft-template}. The template is designed to be modular and flexible, allowing for scalability to accommodate more tasks. Specifically, the template is composed of the following parts: \textcircled{1} targeted binary function (\ie, pseudo code), \textcircled{2} context functions, \textcircled{3} call chains, \textcircled{4} data flow, \textcircled{5} task tag, \textcircled{6} model reasoning process, and \textcircled{7} the final prediction. The \textcircled{1} - \textcircled{5} colored in \textcolor[HTML]{95261F}{red} are assembled to the whole model input, while the \textcircled{6} and \textcircled{7} colored in \textcolor[HTML]{68349A}{purple} are the model output. 
The input includes not only the target function but also the static program analysis results (\ie, \textcircled{2}\textcircled{3}\textcircled{4}), which are considered as the context enhancement being helpful for model analysis, and we will present the details later in \S\ref{sec:context-enhancement}. The ground truth is obtained from source code bridged by debug information (\eg, DWARF); we then reformat them into specific JSON format as the final prediction. However, there is still a critical challenge in building the SFT dataset: the lack of thinking process in the model output. 

Since we aim to build a reasoning model that can take a deep thinking for the user-specific task, we need to provide the model with a large number of examples that contain the reasoning process. It is an intuitive idea to distill existing reasoning LLMs (\eg, DeepSeek-R1) and use reject sampling to collect such dataset. 
Unfortunately, the existing LLMs are not well-trained on binary code and often produce incorrect reasoning processes, leading to mispredictions. Moreover, if one incorporates ground truth into the prompt, the reasoning process of these LLMs tends to refer to them directly, and thus cannot be used as training data. 

\vspace{2ex}
\begin{figure}[htbp]
    \centering
    \includegraphics[width=0.9\textwidth]{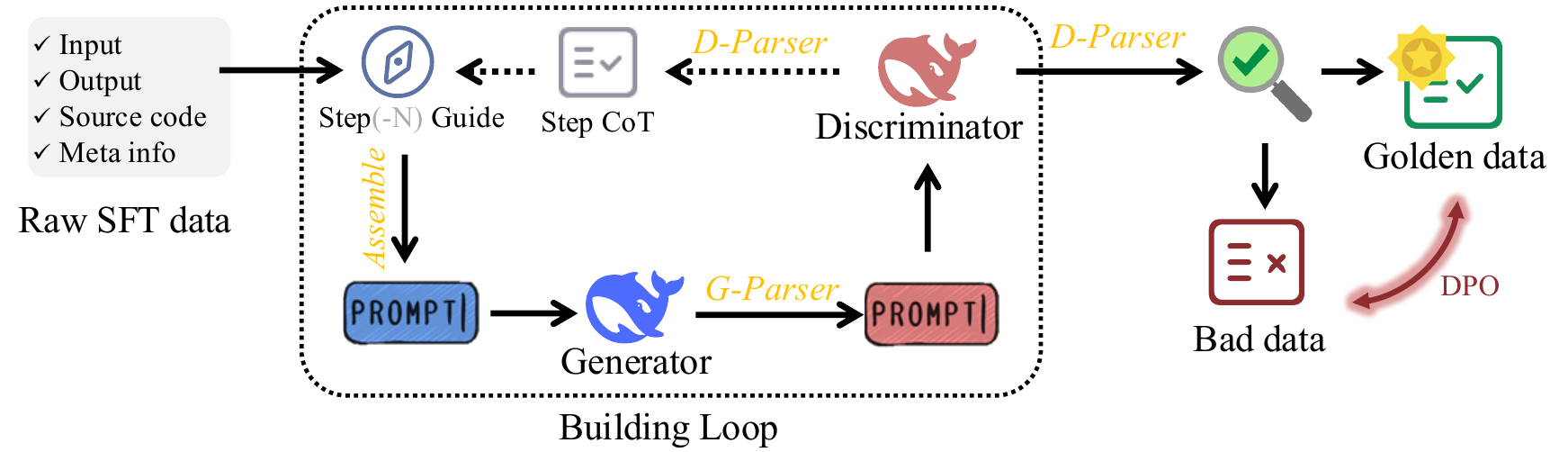}
    \caption{An overview of the generator-discriminator framework for building supervised fine-tuning (SFT) data of binary analysis tasks. The dashed line arrows indicate an optional building loop that generates a single-step chain-of-thought (CoT) each time.}
    \label{fig:generator-discriminator}
\end{figure}

To solve this problem, inspired by previous work in data synthesis~\cite{1billionpersonas-2024, tulu3-2025}, we proposed a generator-discriminator framework to automatically generate SFT dataset with reasoning processes for binary analysis tasks, as shown in \autoref{fig:generator-discriminator}. We first collect the raw SFT data from our raw dataset, which contains the input, output without CoT, corresponding source code, and meta information (\eg, file name and project name). Then, we assemble the raw SFT data and the generation guide into the generator prompt, which is used to prompt a general LLM (\eg, DeepSeek-V3) to generate a CoT without any direct citations to the ground truth. A format-based G-Parser is developed to extract the generated CoT that is subsequently filled into the discriminator prompt. The discriminator is also driven by a general LLM to judge whether the generation meets our requirements. The requirements mainly involve correctness, consistency, helpfulness, and purity. A similar D-Parser is used to access the discriminator's judgment to determinate the destination of the generated CoT. 
In the above program, we invoked the LLM twice to generate one SFT data with a reasoning process, which is efficiently building with a success rate of more than 90\%. However, we found that these CoTs are generally short with an average of 876.33 tokens, especially compared to DeepSeek-R1's thinking. 

\noindent\textbf{Super-CoT.}
To further unleash the power of test-time scaling, we employed a step-by-step building loop to construct super-long chains-of-thought (called Super-CoT), as shown in the dashed box in \autoref{fig:generator-discriminator}. Benefiting from the limitation of our problem scope to the binary domain, we can clearly define the reasoning steps for each domain task and thus perform stepwise generation, which cannot be practiced on general reasoning tasks. In the building loop program, we assemble the raw SFT data and the expert-written step-N guide into the generator prompt to generate the CoT for the current step only. And the discriminator makes a judgment on the current step CoT. If only the current generation is qualified, we will continue to generate the next one with the previous step CoTs integrated. In this way, we invoked the LLM multiple times to generate one SFT data with a Super-CoT, resulting in $\approx$10 times longer reasoning process than before. 
To distinguish it from the usual reasoning process, we modified the thinking tag at the end of the model input to \texttt{<Super-Though>} for enabling the model to conduct Super-CoT.
Nevertheless, this program significantly increases the overhead and reduces the overall success rate, making it costly to scale the Super-CoT dataset.

Beyond the binary analysis SFT dataset, we also sampled general SFT data from open-source datasets, especially the reasoning data in code and math domains. Previous studies~\cite{deepseekr1-2025, mathcoder2-2025-ICLR, Cross-Task-Generalization-2024-naacl} present empirical evidence that reasoning ability is transferable from task to task. For example, the math reasoning ability can be transferred to code reasoning tasks, and vice versa. Finally, we built the SFT dataset with 403K examples, detailed in \autoref{tab:sft-dataset-statistics}.

\begin{table}[htbp]
    \centering
    \caption{Statistics of our supervised fine-tuning (SFT)  dataset.}
    \label{tab:sft-dataset-statistics}
    \begin{tabular}{@{}llrr@{}}
    \toprule
    Data Source                & Category       & \# Samples & \# Tokens(B)    \\ \midrule
    tulu-3-sft-mixture~\cite{tulu3-2025}  &    mixture           & 50,000  & 0.0397           \\
    OpenHermes-2.5~\cite{OpenHermes25-2023}      &   mixture              & 20,000  & 0.0079           \\
    WizardLM\_evol\_instruct\_V2\_196k~\cite{wizardlm-2024} & instruct  & 20,000  & 0.0103           \\
    OpenMathInstruct-2~\cite{openmath2-2024}      &  math           & 10,000  & 0.0049           \\
    OpenO1-SFT~\cite{openo1-2024}                &  reasoning         & 77,685  & 0.0922           \\
    OpenThoughts-114k~\cite{openthoughts114k-2025}   &  reasoning                & 113,957 & 0.7993            \\
    recopilot-sft-cot               & binary        & 99,461  & 0.6088            \\
    recopilot-sft-super-cot         & binary        & 11,781  & 0.1565           \\
    Identity                        & identity        & 465     & 0.0001           \\
    \textbf{In Total}       & \textit{/}  & \textbf{403,349} & \textbf{1.8356}  \\ \bottomrule
    \end{tabular}
\end{table}

\noindent\textbf{DPO dataset.} During the construction of the SFT dataset, low-quality generations were inevitably observed, including issues such as formatting errors and logical inconsistencies. For such cases, we implement a retry mechanism for re-invoking generation with the same raw SFT data. And if the retry produces qualified data, as shown by the highlighted red line in \autoref{fig:generator-discriminator}, the failure-success pair can be utilized as training data for DPO. A DPO training example consists of a chosen response and a rejected response to the same prompt, enabling the model to learn the preference for alignment. Through this approach, we reduced resource wastage by effectively leveraging otherwise unusable data. In total, we have constructed a dataset of 2.4K DPO samples. 

\subsection{Context Enhancement}
\label{sec:context-enhancement}

\begin{figure}[htbp]
    \centering
    \includegraphics[width=0.9\textwidth]{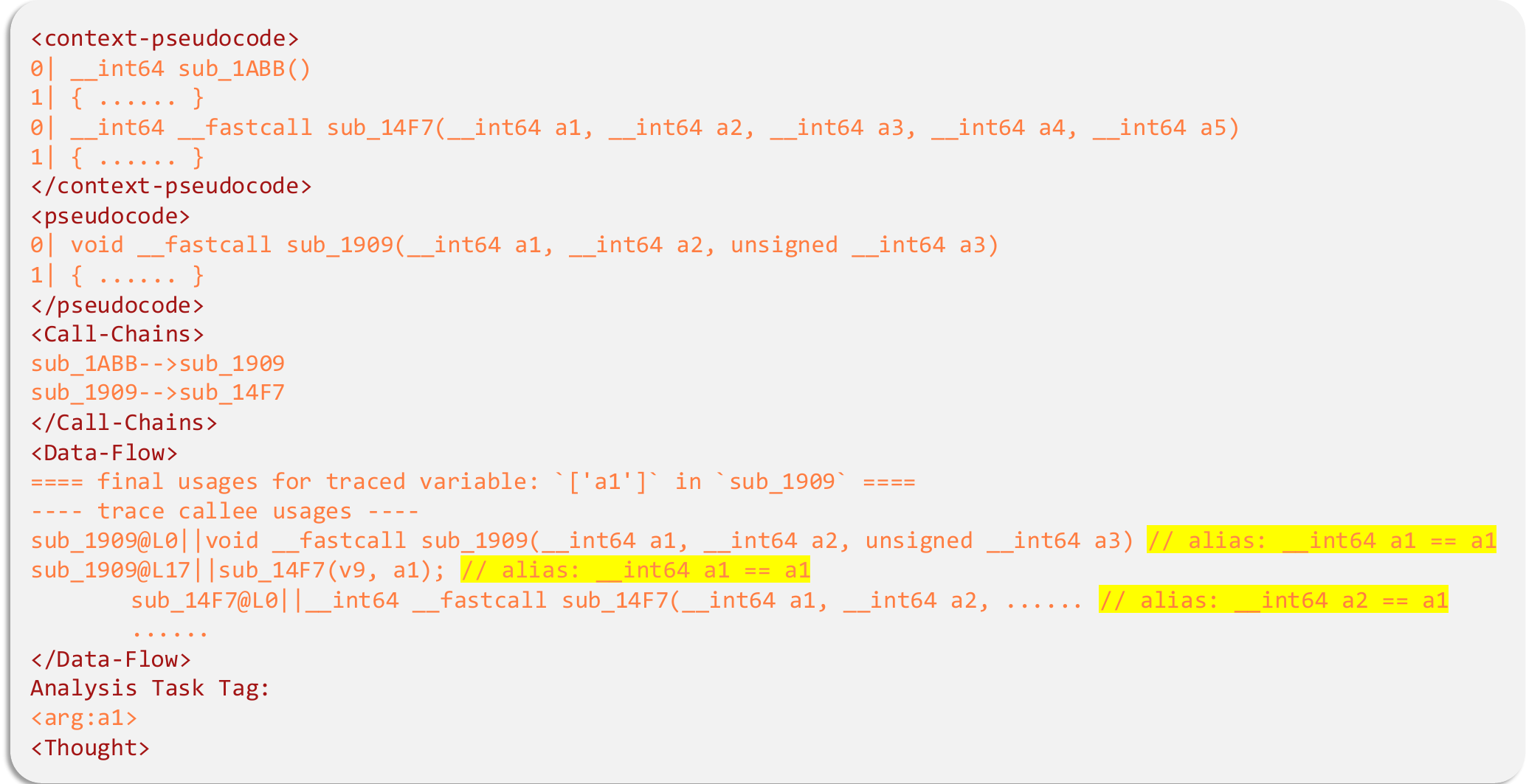}
    \caption{An example of prompt with context enhancement by \sysname. The variable aliases yield from data flow analysis are highlighted in yellow.}
    \label{fig:context-enhancement-example}
\end{figure}

In this section, we present our context enhancement method designed to improve the prompt for analyzing target binary function. To provide a comprehensive view of the binary code, we employ static program analysis on pseudo code to collect context functions, call chains, and variable data flow. Specifically, we first present our call chain analysis (\S\ref{sec:call-chain-analysis}) that traverses the call graph to identify and select the most informative context functions. Then, we introduce our data flow analysis technique (\S\ref{sec:data-flow-analysis}) that traces variable propagation across functions to understand their usage patterns and relationships. As illustrated in \autoref{fig:context-enhancement-example}, these contextual elements are systematically organized in the model input. 

\subsubsection{Call Chain Analysis}
\label{sec:call-chain-analysis}

Starting from the target binary function, we employ a breadth-first search (BFS) strategy to traverse the call graph constructed from decompiled pseudo code. The BFS traversal systematically explores both direct callers and callees of the current function, with the traversal depth limited by a user-specified parameter to prevent context explosion. 
The traversal terminates at leaf nodes (functions with no callees) or root nodes (functions with no callers) in the forward and backward directions respectively, as well as at previously visited functions to avoid cycles. 
Through this bidirectional traversal, we obtain a collection of call chains and their associated context functions. 
We organize the context functions' pseudo code in descending order based on their depth in the call graph, placing functions closer to the target function near the end of the prompt to leverage the model's stronger attention to recent context. 

\noindent\textbf{Informative Score Measurement.} Involving too many functions into prompt can easily exceed the context length limitation. 
To address this challenge, we introduce an empirical metric that quantifies the information richness of a pseudo code function, which correlates with the function's comprehensibility to the LLM. 
This informative score is computed by analyzing three key components: the presence of meaningful function names, the density of string literals, and the semantic information from the callee names:

\begin{equation}
    \mathcal{N}(f) = 
    \begin{cases} 
        1, & \text{if the name symbol of } f \text{ exists } \\
        0, & \text{otherwise}
    \end{cases}
\end{equation}

\begin{equation}
    \mathcal{S}(f) = \mathcal{N}(f) + \min\left(1, \frac{\beta \cdot \text{num\_strings}(f)}{\text{num\_lines}(f)} \right) + \sum\limits_{c \in \text{callees}(f)} \frac{\mathcal{N}(c)}{\lvert \text{callees}(f) \rvert}
\end{equation}

\noindent where $\mathcal{S}(f)$ is the informative score of function $f$, $\mathcal{N}(f)$ is an indicator for meaningful function name, and the $\beta$ is a scaling factor for string density measurement. We empirically set $\beta$ to 25, means the function with 1 string in each 25 pseudo code line is considered as string-rich and informative. 

Finally, we sorted the traversed context functions by their informative scores and selected the top-$k$ functions to be included into the prompt. The parameter of $k$ is configurable according to the model's max context length, which is set to 10 by default in our experiments.
Moreover, in fact, we determine a limited number of context functions not only by the control flow analysis but also by increasing the priority of functions reached by the data flow, which is particularly important for tasks like variable type inference.

\subsubsection{Data Flow Analysis}
\label{sec:data-flow-analysis}

To analyze the source-code type of the variable in binary code, the all usage patterns should be inspected carefully throughout the call graph. We harnessed LLMs to reasoning about it, just like what human experts do. Unfortunately, it is difficult for LLMs to trace a variable's data flow in a bunch of pseudo code, especially when the variable is passed into multiple functions and through a deep call stack. To help LLMs with that, one should build easy-to-read variable usage patterns for LLMs.

We have investigated the existing data flow analysis tools, such as Joern~\cite{joern} and Semgrep~\cite{semgrep}, which could potentially be integrated into \sysname. However, several obstacles prevent their adoption: 1) Although similar, pseudo code is not identical to the C/C++ language, and there is no official support for decompiled pseudo code in these tools; 2) The community version of Semgrep supports intra-procedural data flow only, which is not sufficient for our cross-function analysis needs; 3) Both of them are heavy-weight engines designed for software vulnerability discovery, introducing unnecessary computational overhead in our scenario. Therefore, we opt to implement our own lightweight data flow analysis solution to trace variables in decompiled pseudo code. 

In practice, we trace variable propagations by recursively traversing the abstract syntax tree (AST) of the decompiled pseudo code. Beginning with the target function, we perform a depth-first traversal of the AST. For each variable node encountered, we execute the analysis based on predefined inference rules, detailed in \autoref{fig:inference-rules}.
To avoid too many confusing mathematical symbols, we use programming-like statements in the rules' definition rather than formal Hoare logic notations for better readability. The variable we are interested in is marked as traced at definition statements (\autoref{fig:inference-rules-1}), and will be passed to new variables within the same function by assignment statements (\autoref{fig:inference-rules-3}). We also record each expression using the traced variables (\autoref{fig:inference-rules-2}), which is helpful for the model to understand the variable's usage patterns. When encountering function calls, we recursively traverse the callee's AST following the rules defined for calling statements (\autoref{fig:inference-rules-4}), propagating traced variables through function arguments.
For backward propagation analysis, we employ a breadth-first traversal of the target function's callers, propagating traced arguments into the caller contexts and applying the same analysis to each caller's AST (\autoref{fig:inference-rules-5}). Notably, we no longer handle the function callings in the callers to reduce the overhead.

\begin{figure}[htbp]
    \centering
    \fontsize{8}{9.2}\selectfont
    \begin{minipage}{\textwidth}
        \centering
        \subfloat[Definition Statement]{
            \label{fig:inference-rules-1}
            \begin{minipage}{0.48\textwidth}
                \centering
                \begin{equation}
                    \frac{\texttt { is\_traced(v), is\_def(v) }}{\texttt { set\_alias(v,v), log\_usage(v) }}
                \end{equation}
            \end{minipage}
        }
        \hfill 
        \subfloat[Expression Statement]{
            \label{fig:inference-rules-2}
            \begin{minipage}{0.48\textwidth}
                \centering
                \begin{equation}
                    \frac{\texttt { is\_traced(v), in\_expr(v) }}{\texttt { log\_usage(v) }}
                \end{equation}
            \end{minipage}
        }
    \end{minipage}

    \subfloat[Assignment Statement]{
        \label{fig:inference-rules-3}
        \begin{minipage}{0.9\textwidth}
            \centering
            \begin{minipage}{0.48\textwidth}
                \centering
                \begin{equation}
                    \setlength{\jot}{0pt}
                    \frac{
                        \begin{aligned}
                            & \texttt{is\_traced(v), in\_asg(v), in\_rvalue(v),} \\
                            & \texttt{is\_simple(rvalue), is\_simple(lvalue)}
                        \end{aligned}
                    }{
                        \begin{aligned}
                            & \texttt{set\_alias(lvalue,get\_alias(rvalue)),} \\
                            & \texttt{set\_is\_traced(lvalue), log\_usage(v)}
                        \end{aligned}
                    }
                \end{equation}
            \end{minipage}
            \hfill
            \begin{minipage}{0.48\textwidth}
                \centering
                \begin{equation}
                    \setlength{\jot}{0pt}
                    \frac{
                        \begin{aligned}
                            & \texttt{is\_traced(v), in\_asg(v), in\_lvalue(v),} \\
                            & \texttt{is\_simple(lvalue), is\_simple(rvalue)}
                        \end{aligned}
                    }{
                        \begin{aligned}
                            & \texttt{set\_alias(rvalue,get\_alias(lvalue)),} \\
                            & \texttt{set\_is\_traced(rvalue), log\_usage(v)}
                        \end{aligned}
                    }
                \end{equation}
            \end{minipage}
        \end{minipage}
    }

    \subfloat[Callee Statement]{
        \label{fig:inference-rules-4}
        \begin{minipage}{\textwidth}
            \centering
            \begin{minipage}{0.49\textwidth}
                \centering
                \begin{equation}
                    \setlength{\jot}{0pt}
                    \frac{
                        \begin{aligned}
                            & \texttt {is\_traced(v), in\_callee(v), } \\
                            & \texttt {is\_simple(v), flow\_to(v,callee\_arg) }
                        \end{aligned}
                    }{
                        \begin{aligned}
                            & \texttt { set\_alias(callee\_arg,get\_alias(v)), } \\
                            & \texttt { set\_is\_traced(callee\_arg), log\_usage(v) }
                        \end{aligned}
                    }
                \end{equation}
            \end{minipage}
            \hfill
            \begin{minipage}{0.49\textwidth}
                \centering
                \begin{equation}
                    \setlength{\jot}{0pt}
                    \frac{
                        \begin{aligned}
                            & \texttt { is\_traced(v), in\_callee(v), } \\
                            & \texttt { in\_simple\_expr(v), flow\_to(v,callee\_arg) }
                        \end{aligned}
                    }{
                        \begin{aligned}
                            & \texttt {set\_alias(callee\_arg,refine\_expr(get\_alias(v))),} \\
                            & \texttt {set\_is\_traced(callee\_arg), log\_usage(v) }
                        \end{aligned}
                    }
                \end{equation}
            \end{minipage}
        \end{minipage}
    }

    \begin{minipage}{\textwidth}
        \centering
        \subfloat[Caller Statement]{
            \label{fig:inference-rules-5}
            \begin{minipage}{0.48\textwidth}
                \centering
                \begin{equation}
                    \setlength{\jot}{0pt}
                    \frac{
                        \begin{aligned}
                            & \texttt { is\_traced(v), flow\_to(v,caller\_arg), } \\
                            & \texttt { is\_simple(caller\_arg) }
                        \end{aligned}
                    }{
                        \begin{aligned}
                            & \texttt {set\_alias(caller\_arg,get\_alias(v)),} \\
                            & \texttt {set\_is\_traced(caller\_arg), log\_usage(v) }
                        \end{aligned}
                    }
                \end{equation}
            \end{minipage}
        }
        \hfill 
        \subfloat[Notation Explains]{
            \label{fig:inference-rules-6}
            \begin{minipage}{0.48\textwidth}
                \texttt{is\_simple(x):} \texttt{var |ptr |ref |idx |memptr |memref} \\ 
                \eg, \texttt{(x, *x, \&x, x[y], x.m, x->m)} \\
                \texttt{in\_simple\_expr(x):} \texttt{add |sub} \\ 
                \eg, \texttt{(x-y, x+y)}
            \end{minipage}
        }
    \end{minipage}

    \caption{Inference rules used in data flow analysis for \sysname.}
    \label{fig:inference-rules}
\end{figure}

Throughout the analysis process, we maintain and update alias relationships between the current variables and their original traced variables, indicating how each traced variable is used at different locations.
For example, \autoref{fig:context-enhancement-example} shows a prompt for argument recovery task on \texttt{sub\_1903@a1}, and we traced \texttt{a1} through the call graph to track its propagation and usages. We finally annotated the alias relationship at the variable usage location, \eg, the \texttt{\_\_int64 a2} in \texttt{sub\_14F7} is a direct alias of the original \texttt{a1} in the target function. With these alias annotations, the model is directly aware of the usages of the target variable at all locations that are referring to it without needing to perform analysis itself. 
Our lightweight analysis yields impressive efficiency, specifically tracking all variables for each function takes 0.0182s and 0.0578s for one- and two-level context, respectively.

\section{Benchmark}
\label{sec:benchmark}
\begin{figure}[htbp]
    \centering
    \includegraphics[width=0.95\textwidth]{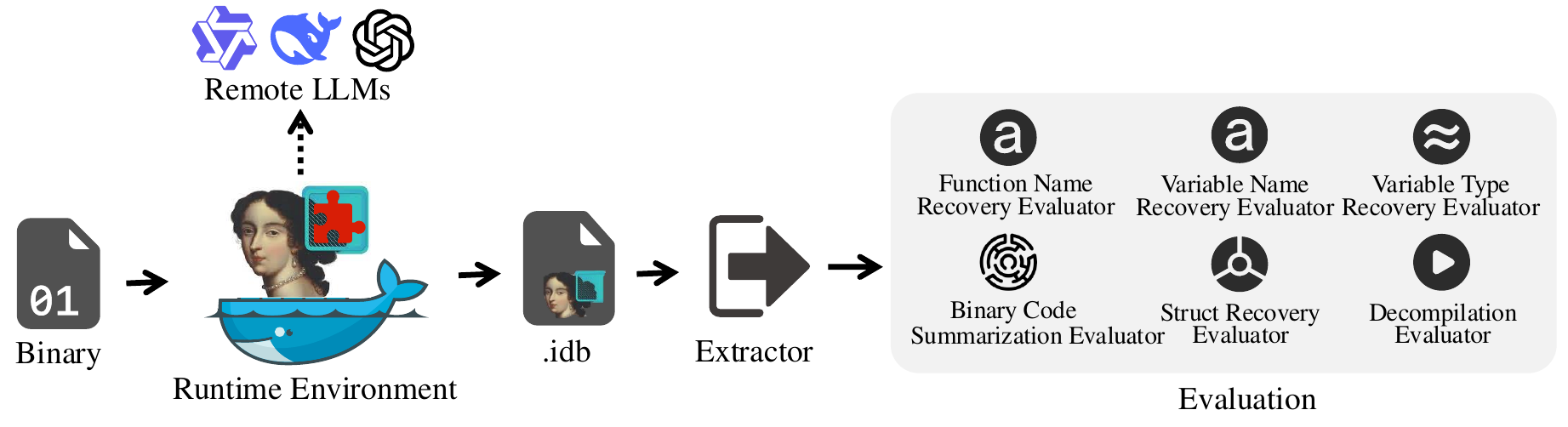}
    \caption{An overview of our benchmark for evaluating binary analysis tools.}
    \label{fig:benchmark-overview}
\end{figure}

As discussed in \S\ref{sec:background}, existing domain benchmarks in binary analysis typically focus on only one or two specific tasks. 
To comprehensively evaluate our \sysname and the baselines, we constructed a multi-task benchmark as shown in \autoref{fig:benchmark-overview}. 
Our goal was to create an automatic and extensible evaluation pipeline for the LLM-based reverse engineering tools. Given that the most popular baselines (detailed in \S\ref{sec:evaluation}) are integrated with IDA Pro~\cite{IDA} as plugins, we employed it to build the runtime environment first, which can be easily extended to other decompilation platforms in the future. 
This pipeline takes a binary as input and triggers a plugin (i.e., a baseline method) to analyze the targeted function at once. 
File-level inputs, rather than function-level used by other benchmarks, provide sufficient context to the potential usage by the evaluation objects. 
Further, the analysis results are persisted by IDA Pro into a \texttt{.idb} file, from which our extractor parses the predictions to run the evaluation.

Our benchmark covers the most important meta-tasks identified from all analysis tasks supported by related works, and we designed corresponding evaluators to provide accurate measurement and valuable insights. These tasks and evaluators are detailed as follows:

\begin{itemize}[left=1em]

\item \noindent\textbf{Function Name Recovery.} We evaluate this task with the Rouge score~\cite{lin-2004-rouge}, which is a recall-oriented metric to indicate how many ground-truth tokens are recalled into the prediction. 

\item \noindent\textbf{Variable Name Recovery.} Since the variable name is similar to the function name, we reuse the same metric.

\item \noindent\textbf{Variable Type Recovery.} In this task, we evaluate only the basic types (\eg, int, float, and char), which appear in different forms in the pseudo code decompiled by IDA Pro. We cluster these equivalent types together, for instance, both \texttt{\_\_int64} and \texttt{unsigned \_\_int64} represent the same type sense with identical memory size. A prediction is considered correct if it belongs to the same cluster as the ground truth. In addition, we ignore the type qualifiers such as \texttt{const}. 

\item \noindent\textbf{Struct Recovery.} Our primary focus here is on recognizing memory patterns of structures, particularly the identification of member numbers and sizes. In practice, understanding memory layouts of complex variables plays a crucial role in enabling further program analysis, such as taint tracking. 
We evaluate each prediction for this task by computing precision and recall of predicted member boundaries, and then derive the F1 score as the final metric:

\vspace{-1ex}
\begin{equation}
\small
\text{Precision} = \frac{|\text{TP}|}{|\text{TP}| + |\text{FP}|}, \quad
\text{Recall} = \frac{|\text{TP}|}{|\text{TP}| + |\text{FN}|}, \quad
\text{F1} = \frac{2 \cdot \text{Precision} \cdot \text{Recall}}{\text{Precision} + \text{Recall}}
\end{equation}
\vspace{-1ex}

\noindent where $TP$ represents correctly predicted member boundaries, $FP$ denotes incorrectly predicted boundaries, and $FN$ indicates ground-truth boundaries that were not predicted.

\item \noindent\textbf{Binary Code Summarization.} Since LLM-as-a-judge had been demonstrated to have a strong alignment with human preference~\cite{llmjudge-2023nips}, with reference to the empirical study~\cite{llmjudgesurvey-2025} and the real-world practice~\cite{opencompass-2023}, we employ a LLM to evaluate generated summarization across the following key dimensions: \textcircled{1} Semantic Coverage, \textcircled{2} Semantic Accuracy, \textcircled{3} Misleading, and \textcircled{4} Readability. 
For each dimension, the LLM makes a binary judgment (yes/no), and the final score is computed as the equally weighted average of these four dimensions, resulting in an overall score between 0 and 1.
For example, if a summarization receives positive judgments in three dimensions but fails in one, it would receive a score of 0.75. \looseness=-1

\item \noindent\textbf{Decompilation.} To evaluate the quality of decompiled code, we employ CodeBLEU~\cite{ren2020codebleu}, a comprehensive metric widely adopted in code synthesis and translation tasks. 
The CodeBLEU score is computed by comparing the tool-generated decompiled code against the original source code across three aspects: syntactic structure (AST), data flow dependencies (DFG), and lexical matching (n-grams).

\end{itemize}

\noindent\textbf{Test Dataset.} The test dataset is built to cover as many domains as possible, specifically, we selected open-source repositories from crypto, network, multimedia, compression, database, sys-utils, etc., and we compiled them with Linux and Windows runtimes. 
Similar to the data collection in \S\ref{sec:raw-data-collect}, we employed sanitizer to filter out noise and prevent data leakage from our training dataset. Although we have no access to the training data of the baseline models, our test dataset is constructed from private environments, suggesting that our data are unlikely to be encountered by them during their training.
In total, we sampled 1,038 binary functions as targets from 60 binaries, which were compiled from 16 projects. 

\noindent\textbf{General Benchmarks.} To further evaluate the general capabilities of our model, we employed a suite of widely recognized benchmarks that target distinct aspects of LLM performance:
\vspace{-1ex}
\begin{itemize}[noitemsep]
    \item \textbf{Mathematical Reasoning:} We use MATH~\cite{letsverify-2023-math500} to evaluate model's proficiency in handling complex mathematical concepts and multi-step reasoning processes.
    \item \textbf{Knowledge and Reasoning:} MMLU~\cite{mmlu-2021} is used to assess the model's knowledge across a wide range of domains, while GPQA-Diamond~\cite{gpqa-2023} specifically assesses the capabilities for expert-level reasoning and understanding in complex questions.
    \item \textbf{Code Generation:} To evaluate the programming abilities across diverse tasks and languages, we utilized HumanEval~\cite{humaneval-chen2021} and MBPP~\cite{mbpp-2021}. These benchmarks focus on the model's capability to generate accurate and functional source code.
    \item \textbf{Instruction Following:} IFEval~\cite{ifeval-2023} is used to measure the model's ability to understand and follow diverse instructions.
\end{itemize}
\vspace{-1ex}
\noindent These additional benchmarks provide a multidimensional view of the model’s general capabilities, allowing us to analyze how domain-specific training influences performance across mathematical reasoning, broad knowledge application, code synthesis, and instruction comprehension. We leveraged the OpenCompass~\cite{opencompass-2023} evaluation platform to systematically benchmark our model alongside baseline LLMs on these tasks.

\section{Evaluation}
\label{sec:evaluation}
\subsection{Experiment Settings}

\noindent\textbf{Training Settings.}
We conducted all experiments on two Linux servers, each equipped with 1TB RAM and 8 * NVIDIA A800-80GB GPUs.
We used \texttt{Qwen2.5-Coder-7B}~\cite{qwen25coder-2024} as the base model and applied our training strategy to produce the \sysname model.
The model checkpoints from each training stage are denoted as \texttt{recopilot-v0.1-beta-*}, with the suffix indicating the stage. For example, \texttt{recopilot-v0.1-beta-dpo} is our final model trained by DPO.
During training, the maximum context length of our model is set to 32K tokens to ensure the effective learning of super-long CoT data. 
We acknowledge the support of open-source training frameworks such as Transformers~\cite{Transformers}, TRL~\cite{trl}, LLaMA-Factory~\cite{llamafactory-2024acl}, and DeepSpeed~\cite{DeepSpeed}, which enable us to conduct efficient training. 
Finally, a full model training goes through CPT, SFT, and DPO stages, and takes about 7.1K GPU hours in total. 

\noindent\textbf{Evaluation Settings.} 
For \sysname, by default, we configure the system with a maximum trace depth of 1 for both callee and caller, a maximum of 10 context functions, and a maximum output length of 16K tokens.
All of the LLM-based methods have inherent randomness, we thus set a maximum of 3 retries for the non-applicable prediction, such as format errors.
Our model uses the prompt template shown in \autoref{fig:sft-template}, which supports all evaluated tasks by specifying task tags, labeled as \texttt{recopilot}. We evaluate baseline methods with the default prompt designed by themselves. 
Our benchmark employed IDA Pro 9.0 as the infrastructure platform in the runtime environment. 

\noindent\textbf{Baselines Selection.} 
We investigated the existing LLM-based tools and methods~\cite{gepetto, wpechatgpt, aidapal, llm4decompile-2024emnlp, mlm01, reverser-ai, ReSym-2024ccs, TypeFSL-2024ase} for the analysis tasks we focused on. The following ones are widely known in the reverse-engineering community, we thus selected them as our baselines:

\begin{itemize}
    \item Driven by General LLM: The Gepetto~\cite{gepetto} and WPeChatGPT~\cite{wpechatgpt} are two open-source tools designed for harnessing general LLMs to analyze binary code, supporting variable name recovery and binary code summarization tasks.
    \item Driven by Tailored LLM: The aidapal~\cite{aidapal} and LLM4Decompile~\cite{llm4decompile-2024emnlp} build their expert LLM for specific tasks. The aidapal supports function name recovery, variable name recovery, and binary code summarization, while the LLM4Decompile supports the decompilation task only. 
    We also include a commercial tool, BinaryNinja~\cite{BinaryNinja}, that provides an AI-powered official extension, \ie, Sidekick. We evaluate it on function name recovery, variable name recovery, and summarization.
\end{itemize}

Beyond these tools, we select state-of-the-art LLMs as the model baselines for model comparison, considering that our \sysname can also be driven by a general-purpose LLM with a tailored prompt.
The \texttt{Qwen2.5-Coder-7B-Instruct}~\cite{qwen25coder-2024} is derived from the same base model as ours, and it demonstrates top-tier performance among models of the 7B parameter size. 
We further take DeepSeek-V3 (671B)~\cite{deepseekv3-2025} and DeepSeek-R1~\cite{deepseekr1-2025} as representative examples of advanced large-scale LLMs and reasoning LLMs, respectively, and compare the performance of our model with theirs.

\subsection{Overall Effectiveness}
\label{sec:overall-effectiveness}

\begin{table}[htbp]
\centering
\small
\setlength{\tabcolsep}{3pt}
\caption{An overall results of comparison against existing tools on binary analysis tasks.}
\label{tab:overall-effectiveness}
\begin{threeparttable}
\begin{tabular}{@{}l|ll|lllllll@{}}
\toprule
Tool                & Model                       & Prompt\tnote{2}    & Succ & Func Name  & Var Name   & Var Type   & Struct      & Dec        & Sum          \\ \midrule
Gepetto             & DeepSeek-V3(671B)           & Self-Def           & \textbf{1.00} & \textit{/} & 7.59       & \textit{/} & \textit{/}  & \textit{/} & 58.26         \\
WPeChatGPT          & DeepSeek-V3(671B)           & Self-Def           & \textbf{1.00} & \textit{/} & 7.79       & \textit{/} & \textit{/}  & \textit{/} & 65.15         \\
aidapal             & aidapal-8k(7B)              & Self-Def           & 0.99 & 9.99       & 9.9       & \textit{/} & \textit{/}  & \textit{/}  & 61.21         \\
LLM4Decompile       & LLM4Decompile-9B-v2         & Self-Def           & \textbf{1.00} & \textit{/} & \textit{/} & \textit{/} & \textit{/}  & 25.51      & \textit{/}   \\
BinaryNinja\tnote{3}        & Sidekick 3.0                  & \textit{/}         & 0.85 & 34.55 & 23.76 & \textit{/} & \textit{/}  & \textit{/}  & 60.85   \\
recopilot-v0.1-beta & recopilot-v0.1-beta-dpo(7B)\tnote{1} & recopilot & 0.92 & \textbf{50.59} & \textbf{43.50}  & \textbf{39.81} & \textbf{27.67}  & \textbf{29.23}  & \textbf{65.21}  \\ \bottomrule
\end{tabular}
\begin{tablenotes}
    \item[1] \texttt{recopilot-v0.1-beta-dpo} is the checkpoint of our expert LLM after DPO training, \ie, the final model.
    \item[2] Self-Def represents self-defined prompts, while recopilot is the prompt template used by our model.
    \item[3] BinaryNinja is a commercial tool that we have no idea about its internal details, and we evaluate it through running the Sidekick extension in batch. 
\end{tablenotes}
\end{threeparttable}
\end{table}

Using our binary analysis benchmark, we conducted experiments to evaluate the performance of our \sysname and the baselines in the domain tasks. We deployed DeepSeek-V3 to power tools designed with general-purpose LLMs, while the tools initiated by tailored models employed their own models. For \sysname, we used the \texttt{recopilot-v0.1-beta-dpo} model with the \texttt{recopilot} prompt and took the configuration by default detailed above. 

As shown in \autoref{tab:overall-effectiveness}, the results demonstrate that \sysname significantly outperforms all baseline methods on almost all tasks, indicating a solid overall effectiveness. 
In particular, \sysname achieves an average outperformance of 13\% over the 2nd place across all tasks. Also, our method implements the functions of variable type recovery and struct recovery that were unsupported by previous LLM-based tools.
Although \sysname currently fulfills the goal of serving as a human assistant, our evaluation results indicate that its absolute performance still has room for improvement before it can be reliably applied to downstream tasks that demand high-level soundness.

As mentioned earlier, the baselines essentially consider a few tasks only, resulting in many blanks in their assessment results (\autoref{tab:overall-effectiveness}), which also suggests their limited value for practical deployment. In contrast, our \sysname model has acquired instruction-following capabilities through the training on both general-purpose and domain-specific tasks, enabling it to perform arbitrary tasks following user prompts. This adaptability broadens its applicability across a wider range of scenarios and enhances its practical significance.

For the particular tasks, we have observed that \sysname achieved the most significant advantage of performance on the variable name recovery task, surpassing the previous best method BinaryNinja by 19.74\%. 
Compared to LLM4Decompile, which is specifically designed for the decompilation task, our method still exhibits a 3.71\% performance advantage.
Regarding binary code summarization, an interesting fact is that the general-purpose LLM also generates summaries well. Specifically, WPeChatGPT and Gepetto with DeepSeek-V3 respectively scored 58.26\% and 65.15\% on this task, which are close to the 65.21\% score of our expert model. It suggests that even non-expert LLMs have shown promising results in assisting binary understanding.

We also count the ratio of LLM generation being successfully applied, denoted as success ratio, presented in the ``Succ'' column of \autoref{tab:overall-effectiveness}. The results indicate that \sysname struggles to generate format-correct and syntactically correct predictions, reaching a 92\% success rate lower than the other methods. Notably, the other methods require only format-agnostic and syntax-agnostic generation, \ie, the model raw output is directly inserted into the placeholders such as comments or function names. However, our model must organize predictions into JSON format, and these predictions may contain errors that make them unusable, such as predicting non-existent variable types.
This weakness indicates a potential direction for \sysname improvement.

\subsection{Model Comparison}
\label{sec:model-comparison}

\begin{table}[htbp]
\centering
\small
\setlength{\tabcolsep}{3pt}
\caption{Comparison of \sysname performance with different models on binary analysis tasks.}
\label{tab:model-comparison}
\begin{threeparttable}
\begin{tabular}{@{}l|llllllll@{}}
\toprule
Model                        & Prompt              & Succ & Func Name & Var Name & Var Type & Struct & Dec    & Sum    \\ \midrule
Qwen2.5-Coder-7B-Instruct    & general             & 0.71 & 34.57     & 24.07    & 30.27    & 9.09   & 23.13  & 50.49   \\
DeepSeek-V3                  & general             & 0.85 & 40.20     & 18.21    & 34.75    & 14.47  & 25.20  & 56.90   \\
DeepSeek-R1                  & general\_wo\_guide  & 0.84 & 43.63     & 25.36    & 38.92    & 22.12  & \textbf{29.77}  & \textbf{66.32}   \\
recopilot-v0.1-beta-Qwen-dpo\tnote{1} & recopilot  & \textbf{0.92} & 46.66     & 32.98    & 37.91    & 26.45  & 26.19  & 62.15   \\
recopilot-v0.1-beta-dpo      & recopilot           & \textbf{0.92} & \textbf{50.59}     & \textbf{43.50}    & \textbf{39.81}    & \textbf{27.67}  & 29.23  & 65.21 \\ \bottomrule
\end{tabular}
\begin{tablenotes}
    \item[1] \texttt{recopilot-v0.1-beta-Qwen-dpo} is a final checkpoint trained without the CPT stage, deriving from the same base model with \texttt{recopilot-v0.1-beta-dpo}.
\end{tablenotes}
\end{threeparttable}
\end{table}

We further conducted experiments to compare the model performance. The different LLMs are used to drive our \sysname tool with the same context enhancement applied, using the default settings. Meanwhile, we designed two prompts, \texttt{general} and \texttt{general\_wo\_guide}, to instruct the non-expert models to generate final results in specific JSON format.
These prompts contain the role description, task statement, formatting instruction, and share the same input specification with the \texttt{recopilot} prompt (shown in \autoref{fig:sft-template}).
The \texttt{general} prompt includes the expert-written stepwise guide for the current analysis task, which is also used in building our SFT dataset (\S\ref{sec:generator-discriminator}), whereas the guide is removed from the \texttt{general\_wo\_guide} prompt. We use the \texttt{general} prompt with regular LLMs (\eg, DeepSeek-V3) for better performance, as these models usually lack domain skills. 
Since the reasoning LLMs (\eg, DeepSeek-R1) have already developed their own reasoning habits, we use another prompt without any guide to set their thoughts free. 

The experimental results are shown in \autoref{tab:model-comparison}.
Among all LLMs, our model, \texttt{recopilot-v0.1-beta-dpo}, achieved the best performance across most tasks, especially showed significant dominance in variable name recovery task, leading the 2nd place DeepSeek-R1 by a 71\% relative score.
According to publicly available leaderboards~\cite{EvalPlus-leaderboard, opencompass-2023}, \texttt{Qwen2.5-Coder-7B-Instruct} is a leading model in the 7B parameter scale.
The average score of our model across all tasks exceeds it by 14\%, indicating a significant performance advantage for our expert model.
Even when compared to advanced large-scale reasoning LLM, DeepSeek-R1, our model still has a small advantage of 5\%. In addition, the smaller size of the \sysname model implies lower resource requirements, overcoming the challenge of laptop-deployability we proposed in \S\ref{sec:background}.
Furthermore, we directly conducted post-training to the base model without domain pretraining, and produced the \texttt{recopilot-v0.1-beta-Qwen-dpo} model. It only achieved sub-optimal performance in the binary analysis tasks compared to the checkpoint with full training performed, indicating that the model can effectively learn domain knowledge through CPT.

\begin{table}[htbp]
\centering
\small
\setlength{\tabcolsep}{3pt}
\caption{Comparison of LLM performance on the benchmarks in the general domains.}
\label{tab:model-comparison-general}
\begin{threeparttable}
\begin{tabular}{@{}l|llllll@{}}
\toprule
Models                       & MATH          & MMLU           & IFEval         & HumanEval          & MBPP          & GPQA-diamond  \\ \midrule
DeepSeek-V3                  & \textbf{81.70} & \textbf{86.71} & \textbf{82.09} & \textbf{91.10}    & \textbf{81.67} & \textbf{56.11}  \\
Qwen2.5-Coder-7b-Instruct    & 43.30         & 65.52           & 58.04          & 84.76             & 73.80          & 31.31          \\
recopilot-v0.1-beta-Qwen-dpo & 37.14         & 63.97          & 57.67          & 82.32             & 69.40         & 34.34 \\
recopilot-v0.1-beta-dpo      & 33.34         & 64.34          & 56.93          & 78.05             & 61.60         & 25.76          \\ \bottomrule
\end{tabular}
\begin{tablenotes}
    \item[*] The scores are practically obtained through our evaluation environment, and minor differences compared to their original reports do not invalidate our comparison of their relative performance.
\end{tablenotes}
\end{threeparttable}
\end{table}

Beyond our domain benchmark, we also employ widely recognized benchmarks to evaluate general capabilities. It is important to note that our primary interest does not lie in the evaluation of the absolute performance. Rather, we aim to provide additional insights by examining how domain-specific training influences the model's general capabilities through comprehensive assessment. We utilize \texttt{Qwen2.5-Coder-7b-Instruct} as a reference baseline to illustrate the impact of our training methodology.

As demonstrated in \autoref{tab:model-comparison-general}, both of the two \sysname models generally underperform compared to the baseline model on general-purpose benchmarks. 
Specifically, \texttt{recopilot-v0.1-beta-dpo}, which went through all three training stages, exhibited the lowest performance, averaging 6.12\% below the baseline across all tasks. This result suggests that our domain-specific training has indeed compromised the model performance in the general domains. 
However, the \texttt{recopilot-v0.1-beta-Qwen-dpo} model, which went through only the SFT and DPO stages, experienced significantly less damage, with an average performance decline of just 1.98\%. 
Notably, we also observe that \texttt{recopilot-v0.1-beta-Qwen-dpo} outperformed the baseline by 3.03\% in the complex reasoning assessment (\ie, GQPA-diamond). This improvement may be attributed to the inclusion of a large amount of general and domain-specific reasoning data in the SFT dataset.
In addition, \texttt{recopilot-v0.1-beta-dpo} consistently underperforms compared to \texttt{recopilot-v0.1-beta-Qwen-dpo} across most benchmarks, with an average performance gap of 4.13\%. This disparity suggests that the CPT training stage may have led to a greater degree of "knowledge forgetting", resulting in the model losing more world knowledge and general capabilities.

\subsection{Ablation Study}
\label{sec:ablation}

\begin{table}[htbp]
\centering
\small
\setlength{\tabcolsep}{3pt}
\caption{Results of ablation experiments on Super-CoT, task guide, DPO training, and context enhancement.}
\label{tab:prompt-ablation}
\begin{tabular}{@{}l|llllllll@{}}
\toprule
Model                   & Prompt                    & Succ & Func Name & Var Name & Var Type & Struct & Dec   & Sum    \\ \midrule
DeepSeek-V3-wo-DFA      & general                   & 0.87 & 35.11     & 17.01    & 34.17    & 9.77   & 24.99 & 59.78  \\
DeepSeek-V3             & general                   & 0.85 & 40.20     & 18.21    & 34.75    & 14.47  & 25.20 & 56.90  \\
DeepSeek-V3             & general\_wo\_guide        & \textbf{0.88} & 35.61     & 18.76    & 30.77    & 14.42  & 24.75 & 60.22  \\
DeepSeek-R1             & general                   & 0.79 & \textbf{46.61}     & 24.87    & 37.76    & 11.38  & \textbf{30.84} & 64.98  \\
DeepSeek-R1             & general\_wo\_guide        & 0.84 & 43.63     & \textbf{25.36}    & \textbf{38.92}    & \textbf{22.12}  & 29.77 & \textbf{66.32}  \\ \midrule
recopilot-v0.1-beta-sft & recopilot                 & 0.82 & \textbf{51.17}   & \textbf{44.01}   & \textbf{40.89}   & \textbf{31.55}  & \textbf{30.11} & \textbf{67.60}  \\
recopilot-v0.1-beta-dpo & recopilot                 & \textbf{0.92} & 50.59     & 43.50    & 39.81    & 27.67  & 29.23 & 65.21  \\
recopilot-v0.1-beta-dpo & recopilot\_super\_thought & \textbf{0.92} & 50.75     & 43.52    & 39.03    & 29.23  & 28.87 & 63.91  \\ 
\bottomrule
\end{tabular}
\end{table}

\noindent\textbf{Super-CoT.} 
We introduced the concept of super long chain-of-thought (Super-CoT) in \S\ref{sec:generator-discriminator}, and used a new thinking tag (\texttt{<Super-Thought>}) to prompt our model to engage in deep and extended reasoning. This prompt is labeled as \texttt{recopilot\_super\_thought} here.
We first conducted an ablation study to evaluate the impact of Super-CoT on the model's performance. 
As shown in \autoref{tab:prompt-ablation}, the performance metrics for \texttt{recopilot-v0.1-beta-dpo} with Super-CoT are similar to the one without it. 
One potential reason for this limited effectiveness is the dataset imbalance. Our Super-CoT dataset comprises only 11K examples, whereas the standard reasoning SFT dataset contains 100K examples. 
This disparity in dataset size may have hindered the model to fully leverage the benefits of deep reasoning, suggesting that expanding the Super-CoT dataset could be a potential direction for our future efforts.
Additionally, we have successfully leveraged SFT to equip the model with basic reasoning capabilities, but it seems to struggle to achieve stable deep reasoning. 
Given the previous studies~\cite{deepseekr1-2025, GRPO-2024}, reinforcement learning is another potential way for improvement.

\noindent\textbf{Task Guide for General LLM.} 
We further conducted an evaluation to investigate the \texttt{general} prompts, revealing that general LLMs like DeepSeek-V3 benefit from task-specific guides. With guided prompts, DeepSeek-V3 achieved a higher average score of 31.62\%, compared to 30.60\% without guidance, indicating that guided prompts help the general-purpose model skilled binary analysis tasks.
Conversely, for reasoning models like DeepSeek-R1, removing the guide improved performance, with an average score increase from 36.01\% to 37.68\%. This indicates that such models perform better when allowed to conduct their inherent reasoning capabilities without constraints.

\noindent\textbf{DPO.} 
We evaluated the \texttt{recopilot-v0.1-beta-sft} model, a prior checkpoint of \texttt{recopilot-v0.1-beta-dpo}, which had not yet undergone DPO training. By comparing the evaluation results of these two models, we observed that DPO training significantly enhances the success ratio, increasing it from 0.82 to 0.92. This improvement underscores the effectiveness of DPO training in refining the model's ability to produce format-correct and syntactically accurate predictions, which are essential for the integration into decompiled code.
While the success rate has improved, there are declines in certain tasks, such as struct recovery (3.87\%) and decompilation (0.87\%). This suggests that while DPO optimizes the model's prediction reliability, it may slightly compromise its performance in specific tasks. These findings highlight the importance of balancing the preference alignment and the original performance.


\noindent\textbf{Data Flow Analysis.}
In order to evaluate the impact of data flow analysis (DFA), as shown in \autoref{tab:prompt-ablation}, we conduct a comparison and present results in the rows of \texttt{DeepSeek-V3} and \texttt{DeepSeek-V3-wo-DFA}. 
Our \sysname model is trained with data that includes DFA, performing ablation on them would naturally lead to performance degradation due to missing information.
Therefore, we chose to conduct this ablation experiment on the DeepSeek-V3 model to isolate and assess the specific contribution of DFA to the analysis performance of LLMs.
The evaluation results demonstrate that the inclusion of DFA significantly improved performance in specific tasks. For instance, DeepSeek-V3 with DFA exhibits 4.70\% higher accuracy in identifying struct layouts, highlighting the importance of understanding variable propagation and usage patterns. 
This improvement underscores the value of DFA in providing the model with a more comprehensive understanding of the code's data flow, thereby facilitating more accurate and insightful analysis.

\section{Discussion}
\label{sec:discussion}

In the previous section, the evaluation demonstrates that our method outperforms the baseline methods. 
Although the absolute performance of \sysname still needs to be further improved, it has shown the potential of becoming the next generation of decompilation assistants.
This section will discuss the limitations of our work and potential ways for future improvements.
Starting with the weaknesses of \sysname in practice:



\noindent\textbf{Weakness on Supported Tasks.}
Binary analysis is a complex engineering that involves different analytical tasks depending on the objectives. 
For instance, malware analysis often requires code clustering to identify functional modules, while deobfuscation may necessitate control flow optimization. Constrained by computation resource, \sysname currently focuses primarily on the most fundamental and critical tasks for decompilation augmentation, which leads us to temporarily overlook higher-level and peripheral requirements.

\noindent\textbf{Weakness on Supported Binary Representations.}
\sysname currently works on decompiled pseudo code only, which leads to weaknesses in handling other representations of binary code, such as disassembly code. In practice, some binaries for specialized CPU architectures lack a decompiler to obtain pseudo code. We plan to extend our model's ability in tackling disassembly code in the future, thereby supporting more practical scenarios.

\noindent\textbf{Weakness on Supported Programming Languages.}
Recently, a growing number of compiled high-level languages have been widely used, such as Go and Rust. However, our training data are exclusively coming from C/C++ projects, which prevents our model from learning binaries built using other programming languages. The baseline methods suffer from the same problem, underscoring a pressing need for dataset expansion.


The weaknesses presented above could be mitigated by taking more efforts on dataset building and model training on our current method, that is, there are no serious obstacles.
Furthermore, we identified several potential approaches to improve the performance of our model in the future:

\noindent\textbf{Reinforcement Learning.} 
Reinforcement learning in the LLM domain has shown significant efficacy in improving model performance~\cite{RLHF-2022, GRPO-2024, deepseekr1-2025}. 
While our model has acquired a certain level of reasoning ability through SFT directly, it struggles to perform stable and consistent logical reasoning in a super long CoT, detailed in the ablation study (\S\ref{sec:ablation}). 
However, the on-policy reinforcement learning, which continuously optimizes the model starting from itself, could help the model naturally develop robust ability of test-time scaling.

\noindent\textbf{Model and Dataset Scaling.}
The neural scaling law~\cite{scalinglaw-2020} has been empirically validated, stating that model performance improves as both model size and dataset size increase. Our current 7B model is designed to be deployable on personal laptop, which inherently limits its upper bound. A promising direction for optimization is to scale up the model parameters, as well as the diversity of data and the number of domain tasks.

\noindent\textbf{Agentic Mode.}
LLM-driven agents have achieved promising results in automating sophisticated tasks, as exemplified by tools like Deep Research~\cite{deepresearch-openai} and Cursor~\cite{cursor}. Unlike traditional LLMs, which primarily engage in Q\&A interactions, agents are capable of autonomous planning to solve problems through multiple steps and invoke external tools to obtain additional information or assistance. Building agentic capabilities on top of \sysname model presents a promising approach to addressing more complex binary analysis challenges.

\section{Conclusion}
\label{sec:conclusion}

In this work, we presented \sysname, an expert LLM, tailored to provide assistance with reverse engineers in binary analysis.
To build the model, we took efforts to collect large-scale raw dataset and devised a generator-discriminator framework to construct CoT data. 
We further employed context enhancement through data flow and call graph analysis for better performance. 
A benchmark has been implemented for binary analysis with the most important tasks supported.
Our comprehensive evaluation showed that \sysname outperforms existing domain-specific LLMs and advanced general LLMs.
We have elaborated on the implementation details of \sysname and demonstrated this work to the security community. 
We hope this work promotes the security community to drive binary reverse engineering into the next generation.

\bibliographystyle{unsrturl}
\bibliography{paper.bib}

\end{document}